\documentclass[prb,aps,twocolumn,amsmath,amssymb,floatfix,
superscriptaddress]{revtex4}
\usepackage[dvips]{graphics}
\usepackage{color}
\definecolor{dred}{rgb}{0,0,0.6}
\usepackage{graphicx}  
\usepackage{dcolumn}   
\usepackage{bm}        
\usepackage{amssymb}   
\usepackage{amsmath}
\usepackage{braket}
\usepackage{upgreek}
\usepackage[mathscr]{euscript}
\usepackage{color}
\usepackage[toc,page]{appendix}
\usepackage[colorlinks=true, letterpaper=true, pdfstartview=FitV, linkcolor=blue, citecolor=blue, urlcolor=blue]{hyperref}

\begin{document}

\title{Thermolectricity in irradiated bilayer graphene flakes}

\author{Cynthia Ihuoma Osuala}
\email{cosuala@stevens.edu}
\affiliation{Department of Physics, Stevens Institute of Technology, Hoboken, NJ 07030, USA}

\author{Tanu Choudhary}
\email{tanunain27@gmail.com}
\affiliation{Department of Physics, Faculty of Natural Sciences, M. S. Ramaiah University of Applied Sciences, Bengaluru 560058, India}

\author{Raju K. Biswas}
\email{rajukumar1718@gmail.com}
\affiliation{Department of Physics, Faculty of Natural Sciences, M. S. Ramaiah University of Applied Sciences, Bengaluru 560058, India}

\author{Sudin Ganguly}
\email{sudinganguly@gmail.com}
\affiliation{Department of Physics, School of Applied Sciences, University of Science and Technology Meghalaya, Ri-Bhoi 793101, India}

\author{Santanu K. Maiti}
\email{santanu.maiti@isical.ac.in}
\affiliation{Physics and Applied Mathematics Unit, Indian Statistical
  Institute, 203 Barrackpore Trunk Road, Kolkata 700108, India}

\date{\today}

\begin{abstract}
We present a comprehensive study on enhancing the thermoelectric (TE) performance of bilayer graphene (BLG) through irradiation with arbitrarily polarized light, focusing on $AA$- and $AB$-stacked configurations with zigzag edges. Utilizing a combination of tight-binding theory and density functional theory (DFT), we systematically analyze the impact of light irradiation on electronic and phononic transport properties. Light irradiation alters the electronic hopping parameters, creating an asymmetric transmission function, which significantly increases the Seebeck coefficient, thereby boosting the overall {\it figure of merit} (FOM). For the phononic contribution, DFT calculations reveal that $AB$-stacked BLG exhibits lower lattice thermal conductivity compared to $AA$-stacked, attributed to enhanced anharmonic scattering and phonon group velocity. The combined analysis shows that FOM exceeds unity in both stacking types, with notable improvements near the irradiation-induced gap. Additionally, we explore the dependence of FOM on the system dimensions and temperature, demonstrating that light-irradiated BLG holds great promise for efficient thermoelectric energy conversion and waste heat recovery. Our results show favorable responses over a wide range of irradiation parameters. These findings provide crucial insights into optimizing BLG for advanced TE applications through light-induced modifications.
\end{abstract}

\maketitle
\section{\label{sec1} Introduction}

Thermoelectric materials which convert thermal energy into electrical energy and vice versa, hold significant promise for sustainable energy technologies and waste heat recovery~\cite{Xu2014,Amollo2018}. The efficiency of these materials is quantified by the dimensionless {\it figure of merit} (FOM) $ZT$, which depends on the Seebeck coefficient, electrical conductivity, and thermal conductivity. Materials with a 
$ZT<1$ are usually deemed inefficient for real world uses whereas those with a $ZT>1$ are seen as having strong thermoelectric performance. The goal generally is to reach a $ZT>2$ for commercially feasible solutions~\cite{tritt}. This has spurred significant research efforts into enhancing thermoelectric efficiency through new materials and structural innovations, as achieving high $ZT$ can lead to better performance in power generation and refrigeration technologies.

In recent years, significant advancements in the TE sector have revealed that low-dimensional materials, such as two-dimensional (2D) materials and nanostructures, exhibit superior performance compared to their bulk counterparts~\cite{Hicks,Hicks1993,Dresselhaus2007}. These low-dimensional systems benefit from quantum confinement effects and enhanced boundary scattering, which significantly influence the transport properties of electrons and phonons. Quantum confinement can improve the Seebeck coefficient by discretizing energy levels, thereby sharpening the density of states at the Fermi level and increasing voltage generation from a given temperature gradient~\cite{Je2009, Oxandale2023}. Additionally, boundary scattering reduces thermal conductivity by limiting the mean free path of phonons, minimizing heat conduction~\cite{Regner2013}. These combined effects result in a higher FOM, making low-dimensional materials more suitable for efficient thermoelectric devices. However, the technology to obtain such structures is often complex and expensive, posing challenges for commercialization. Despite these challenges, extensive research continues to focus on developing nanostructures and other low-dimensional systems for advanced thermoelectric applications.

Graphene, a single layer of carbon atoms arranged in a hexagonal lattice, is widely known for its exceptional electrical, mechanical, and thermal properties, such as its flexibility~\cite{Kim2009, Jang2016}, high carrier mobility~\cite{Bolotin2008, Banszerus2015,Banszerus2016}, and thermal conductivity~\cite{Balandin2008}. These attributes make it an excellent candidate for applications requiring efficient heat dissipation. However, the extremely high thermal conductivity of graphene poses a significant challenge for its use in thermoelectric applications~\cite{Shih2013}, as it results in a low overall $ZT$ value. 
Bilayer graphene (BLG), which consists of two stacked layers of graphene~\cite{McCann2013}, retains many of the desirable properties of single-layer graphene while offering additional tunability through interlayer interactions and stacking order. BLG is a vertical stacking of hexagonal layer along their $c$-axis, coupled
by the van der Walls interaction, with two primary stacking configurations, namely $AA$-stacked and
$AB$- stacked BLG. In general, natural graphite adopts the $AB$-stacking sequence because a
small energy barrier exists between the $AB$- and $AA$-stacking ones. Moreover, recently, Liu et
al. found that the BLG often exhibits $AA$ stacking~\cite{liu-prl}, which is very hard to distinguish from the monolayer graphene. However, Kong et al.~\cite{kong} reported that transitioning from $AA$ to $AB$ stacking
has no significant effect on the qualitative properties of lattice thermal conductivity, only the
magnitude is reduced slightly by about 100$\,$W/mK. Interestingly, BLG exhibits a tunable bandgap, which is advantageous for electronic and optoelectronic applications, allowing for greater control over the electronic properties for various applications~\cite{Zhang2009, Choi2010, Ramasubramaniam2011}. The interlayer coupling in BLG can be adjusted by external electric fields, chemical doping, or strain, providing a means to modulate its electronic properties in ways that are not possible with single-layer graphene. Despite these  promising features, BLG still faces challenges similar to those of single-layer graphene, particularly its high thermal conductivity, which limits its thermoelectric efficiency.
This property, while beneficial for heat dissipation, limits the thermoelectric efficiency of BLG by preventing the maintenance of a sufficient temperature gradient necessary for efficient thermoelectric conversion. The high thermal conductivity leads to rapid heat dissipation, which counteracts the development of a thermal gradient that drives thermoelectric processes. Therefore, a critical challenge in utilizing BLG for thermoelectric applications is to reduce its thermal conductivity while preserving its excellent electrical properties.

To address these challenges, several studies have investigated various methods to reduce the thermal conductivity of graphene through techniques such as the introduction of domain boundaries~\cite{Yasaei2015}, carbon isotopes~\cite{Chen2012,Anno2014,Fthenakis2014},  structural defects~\cite{Xie2014,Fthenakis2014,Hao2011},  wrinkles~\cite{Chen2012,Wang2014}, and functionalization~\cite{Mu2014}.
These approaches aim to disrupt the phonon transport pathways, thereby reducing thermal conductivity and improving the 
$ZT$ value. However, each method comes with its own set of trade-offs, such as potential impacts on electrical conductivity and mechanical integrity.

In this work, {\it we propose an alternative approach to enhancing the TE response of BLG via light irradiation}. This process selectively modifies the directional electronic hopping parameters~\cite{gomez-prl}, resulting in an asymmetric electronic transmission function. An asymmetric transmission function is a key factor in improving the TE response~\cite{mahan}. Building on this promising approach, we examine two primary stacking configurations of BLG: $AA$-stacked and $AB$-stacked, subjected to irradiation with arbitrarily polarized light. The impact of light irradiation is incorporated through a tight-binding model to analyze the electronic transport properties. To comprehensively understand the phonon transport characteristics, we employ {\it ab-initio} density functional theory (DFT) calculations combined with Boltzmann transport theory. This study provides a detailed analysis of the electron and phonon spectra, transmission function, electrical conductance, thermopower, thermal conductance due to both electrons and phonons, atomic vibration modes, phonon group velocity, and Gr\"{u}neisen parameters, offering a thorough understanding of the heat transport properties.

The remainder of the paper is organized as follows: In the next section (Sec.~\ref{sec2}), we discuss the structural properties of BLG, including the electronic band structure, the tight-binding Hamiltonian, the theoretical framework for light irradiation, and the equations used to compute various electronic thermoelectric quantities. Following this, we present our results in Sec.~\ref{sec3}, detailing the electronic and phononic transport properties. Finally, we conclude in Section~\ref{sec4}.

\section{\label{sec2} Theoretical formulation}

\subsection{Thermoelectric setup and tight-binding Hamiltonian}
\begin{figure}[h]
\centering
\includegraphics[width=0.49\textwidth]{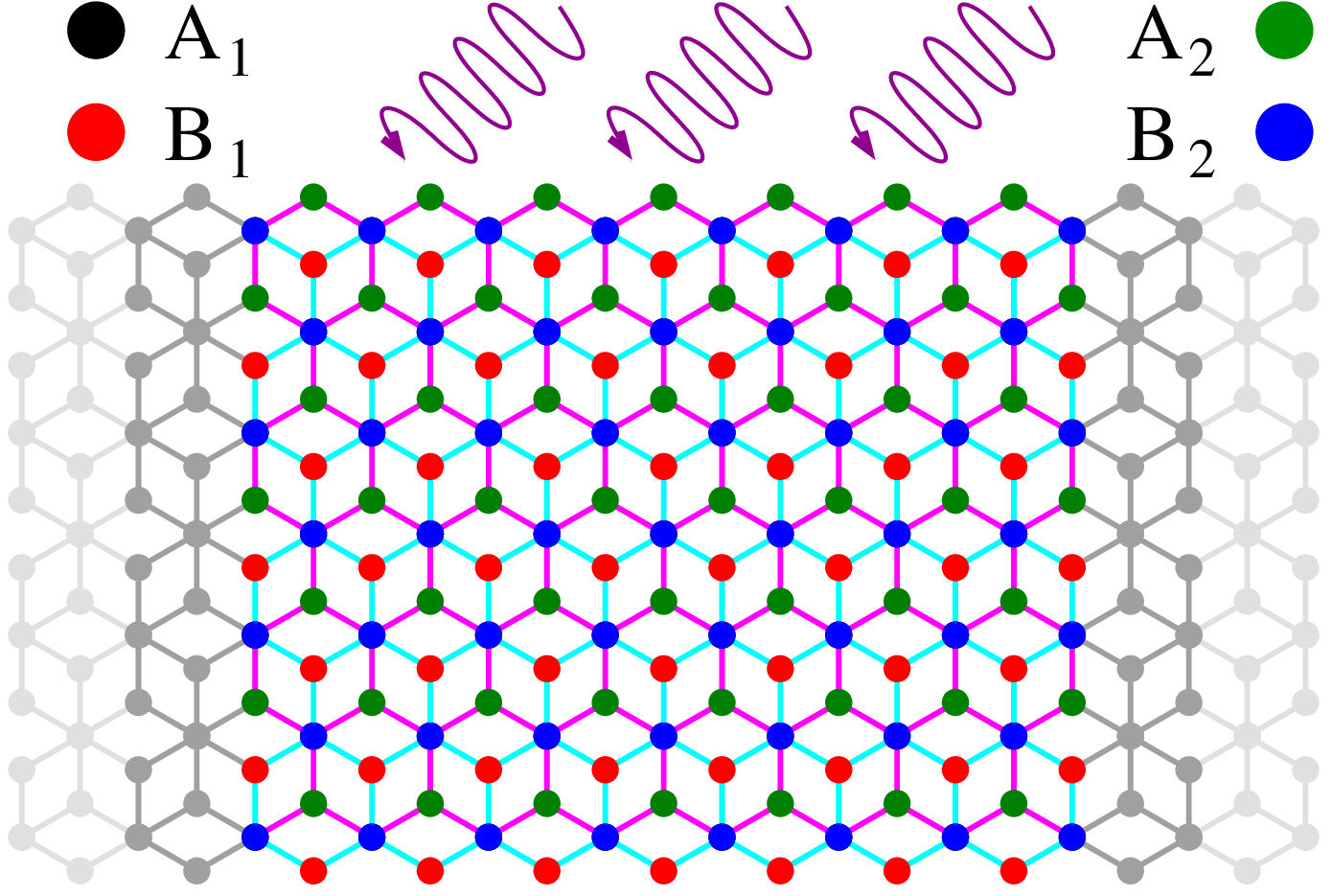}
\caption{(Color online). Schematic lattice structure of bilayer graphene nanoribbon with $AB$ stacking which is connected in between two semi-infinite electrodes, namely source and drain (shown by grey clolor), kept at temperatures, $T + \Delta T/2$ and $T - \Delta T/2$, respectively, where $\Delta T$ is an infinitesimally small temperature difference. Black and red circles represent atoms $A_1$ and $B_1$ on the bottom layer, while the upper layer is represented by green and blue circles corresponding to $A_2$ and $B_2$ respectively. $A_1$ atoms (black circles) are not visible in the figure as they are overlapped by $B_2$ atoms (blue circles). The bonds in the bottom layer is denoted with cyan and those in the top layer by magenta. The purple waves represent the incident light irradiation perpendicular to the surface.}
\label{sys}
\end{figure}

Figure~\ref{sys} depicts the schematic diagram of a zigzag bilayer graphene flake, modeled as a two-coupled monolayer of carbon atoms, each with a honeycomb crystal structure. To evaluate the transmission probability and explore the performance of the thermoelectric effect, we symmetrically 
attach two semi-infinite electrodes (depicted in grey) to each end of the graphene flakes. These electrodes having the same lattice structure as that of the flake are connected through the same layer of the bilayer region and are maintained at two different temperatures, denoted as $T+\Delta T/2$ and $T-\Delta T/2$, respectively, where $\Delta T$ is an infinitesimally small temperature. This allows us to confine our analysis to the linear response regime.

We describe our system using a spinless tight-binding framework which can potentially capture the essential physics of quantum transport ~\cite{neto, mccann, jung}.
The Hamiltonian of the central AB-stacked flake is expressed as~\cite{jwgon}
\begin{eqnarray}
H_{AB} &=& \sum_{\langle nm\rangle, \alpha}\left(
t_{nm}a_{\alpha,n}^{\dag}b_{\alpha,m} + \text{h.c.}\right) \nonumber\\
&+& \gamma_0\sum_{m}  \left(
a_{1,m}^{\dag}b_{2,m} + \text{h.c.}\right),
\label{ham}
\end{eqnarray}
where $a_{\alpha,n}\left(a_{\alpha,n}^\dag\right)$ and $b_{\alpha,n}\left(b_{\alpha,n}^\dag\right)$ denote the fermionic annihilation (creation) operators on sublattices $A$ and $B$, respectively, in the plane $\alpha=1,2$, corresponding to the lattice site $n$. The angular brackets represent in-plane nearest-neighbor bonds. $t_{nm}$ is the nearest-neighbor hopping strength. The second term in Eq.~\ref{ham} denotes the coupling between the two layers and $\gamma_0$ is the corresponding hopping strength. 
The electrodes are also described within the nearest-neighbor tight-binding framework.

\subsection{Incorporation of light irradiation}
When light irradiates a system, it transforms the system into a periodically driven one, thereby introducing complexity and challenges. However, in the minimal coupling regime, addressing such a time-dependent issue can be simplified by employing the Floquet-Bloch ansatz~\cite{gomez-prl, sambe, delplace, martinez}. Following the Floquet approximation, the influence of light incorporation can be managed through a vector potential 
$\mathbf{A}_{ac}(\tau)$. Within the tight-binding framework, the vector potential is incorporated into the hopping integral using the Peierls substitution $\frac{e}{c\hbar}\int {\mathbf{A}}_{ac}(\tau)\cdot d{\mathbf{l}}$, with the symbols $e$, $c$, and $\hbar$, representing physical quantities. Without loss of generality,  the vector potential can be expressed as follows: ${\mathbf{A}}_{ac}(\tau) = (A_{x}\sin(\omega\tau),  A_{y}\sin(\omega\tau + \varphi),  0)$, signifying an arbitrarily polarized field in the X-Y plane. Here, $A_x$ and $A_y$ denote the field amplitudes, and $\varphi$ represents the phase. Following rigorous mathematical calculations, the effective hopping integral under irradiation is expressed as~\cite{gomez-prl,delplace}
\begin{equation}
t_{nm} \rightarrow t_{nm}^{pq} = t_{nm}\times\frac{1}{\mathbb{T}}\int_0^{\mathbb{T}}e^{i\omega\tau(p-q)}e^{i\mathbf{A}_{ac}(\tau)\cdot \mathbf{d}_{nm}}d\tau, 
\label{effhop}
\end{equation}
where $\mathbf{d}_{nm}$ denotes the vector connecting nearest-neighbor sites in the BLG and $t_{nm}^{pq}$ is the modified hopping integral due to irradiation. $p$ and $q$ are associated with the Floquet bands. $t_{nm}$ is assumed to be isotropic and represented as $t$. 

Assuming a uniform driving field characterized by frequency $\omega$ and time-period $\mathbb{T}=2\pi/\omega$, the vector potential is quantified in units of $ea/c\hbar$, with $a$ denoting the lattice constant. By considering the explicit form of $\mathbf{A}_{ac}(\tau)$, Eq.~\ref{effhop} can further be expressed as
\begin{equation}
t_{nm}^{pq} = t \mathrm{e}^{i(p-q)\Theta} J_{(p-q)}(\Gamma),
\label{hop-light}
\end{equation}
where $J_{(p-q)}$ denotes the $(p-q)$-th order Bessel function of the first kind~\cite{delplace, martinez}. The terms $\Gamma$ and $\Theta$ are
\begin{eqnarray}
\Gamma &=& \sqrt{(A_{x}d_{x})^2 + (A_{y}d_{y})^2 + 2A_{x}A_{y}d_{x}d_{y}\cos\varphi}\,,  \\
\Theta &=& \arctan \left(\frac{A_{y}d_{y}\sin\varphi}{A_{x}d_{x}+A_{y}d_{y}\cos\varphi}\right).
\end{eqnarray}

The layers are considered to lie in the $X-Y$ plane, resulting in the coupling between the two layers occurring along the $Z$-direction. Consequently, the corresponding bond is purely oriented in the $Z$-direction. Based on the above analysis, it is clear that the vertical hopping parameter, $\gamma_0$, remains unaffected by the irradiation. However, since the in-plane bonds are associated with different vectors, light irradiation induces anisotropy in the hopping strengths corresponding to different bond directions. For a detailed mathematical description, please refer to Refs. \cite{gomez-prl, delplace}.

\subsection{Computation of electronic thermoelectric quantities}
\subsection*{Transmission probability}
We utilize the KWANT~\cite{groth} package to evaluate the two-terminal transmission probability of our system connected to finite-width electrodes. KWANT streamlines this assessment through the wave function approach within the scattering matrix formalism. This technique, equivalent to the non-equilibrium Green's function method with the Fisher-Lee relation~\cite{etms,qtat}, incorporates the influence of electrodes through self-energy terms, treated as a superposition of plane waves. After determining propagating modes, it (KWANT) solves the tight-binding equations $H\psi_i = \epsilon\psi_i$ to derive the scattering matrix. Here, $\psi_i$ represents the wave function confined to the scattering region, and $H$ is the total Hamiltonian in tridiagonal block form, encompassing the Hamiltonian matrix of the scattering region, the Hamiltonian related to the unit cell of the electrodes, and other relevant block matrices.

\subsection*{Thermoelectric quantities}
Once we compute the two-terminal transmission probability, we can determine various thermoelectric quantities, which in turn allow us to calculate $ZT$. This dimensionless parameter measures the efficiency of a thermoelectric material and is given by the formula
\begin{equation}
    ZT=\frac{GS^{2}T}{k}.
\end{equation}
Here, $G$ represents the electrical conductance, $S$ is the Seebeck coefficient (or thermopower), and $k(=k_{el}+k_{ph})$ is the total thermal conductance, including contributions from both electrons $(k_{el})$ and phonons $(k_{ph})$. $T$ represents the equilibrium temperature of the central system. In the linear regime, each of these quantities, except $(k_{ph})$, can be calculated using the Landauer integral~\cite{finch,zerah, kallol-scirep} as
\begin{subequations}
\begin{align}
G&=\frac{2e^2}{h}L_0, 
\label{eq:10a}\\
S&=-\frac{1}{eT}\frac{L_1}{L_0},
\label{eq:10b}\\
k_{el}&=\frac{2}{hT}\bigg(L_{2}-\frac{L_1^2}{L_{0}}\bigg)
\label{eq:10c}
\end{align}
\label{eqn:2}
\end{subequations}

The Landauer integral, $L_n$ is defined as
\begin{equation} L_n=-\int {\mathcal{T}}(E)(E-E_F)^n \left(\frac{\partial f(E)}{\partial E}\right)dE.
 \label{eqn:11}
 \end{equation}
Here, ${\mathcal{T}}(E)$, $E_F$, and $f(E)$ represent the two-terminal transmission probability, Fermi energy, and Fermi-Dirac distribution function, respectively. 

To obtain an accurate value of $ZT$, it is crucial to account for the thermal conductance due to phonons, especially at finite temperatures. We study this using density functional theory, which is discussed as below.

The first-principles based DFT simulations are implemented in Quantum Espresso (QE) package~\cite{dft}. The simulations use pseudopotentials to accurately account for electron-core interactions and effectively describe the behaviour of valence electrons in the crystal. We utilize ultrasoft pseudopotentials in the local density approximation with Perdew, Burke, and Ernzerhof (PBE)~\cite{pbe} functional to obtain the structural and phonon transport properties of the $AA$- and $AB$-stacked BLGs. In the optimizing simulations, energy cutoff of 200 $Ry$ and 50 $Ry$ are used for charge density and wave functions. The van der Waals interaction is taken into account by the Grimme D3~\cite{d3}, which plays an important role in bilayer graphene. 

The lattice thermal conductivity is calculated using the ShengBTE package~\cite{bte} with an iterative self-consistent technique, which involves solving the phonon Boltzmann transport equation. The QE code~\cite{dft} is employed to obtain the second-order inter-atomic force constants (IFCs). We obtain the phonon dispersion with a $q$ mesh of 2$\times$2$\times$2 with a strict convergence threshold in the self-consistent (scf) calculation using density functional perturbation theory (DFPT)~\cite{dfpt}. An 8$\times$8$\times$2 Monkhorst-Pack $k$-mesh is used in the self-consistent field (SCF) calculation. The third-order force constants are determined through a finite displacement approach, wherein three atoms in the supercell are simultaneously displaced from their equilibrium position, and the resulting forces on the remaining atoms are computed. A 3$\times$3$\times$2 supercell is constructed, allowing interactions with three nearest-neighboring atoms for the calculation of third-order IFCs.

\section{\label{sec3}Results and Discussion}
Before we present our findings, it is important to mention the frequency of the incident light. When a system is illuminated with light irradiation, corresponding Hamiltonian becomes time-dependent. However, such a time-dependent ${\mathbb D}$-dimensional lattice can be described as a $({\mathbb D}+1)$-dimensional static lattice~\cite{gomez-prl, delplace}, where the additional dimension comes from the time-periodicity. In this framework, the original Bloch band splits into multiple Floquet-Bloch (FB) bands, with the coupling between these bands dependent on the driving frequency. The $({\mathbb D}+1)$-dimensional lattice can be envisioned as a series of virtual replicas of the original lattice stacked vertically~\cite{gomez-prl}.

In the low-frequency regime, the coupling between the parent lattice and its virtual copies is significant, and higher-order terms in $p$ and $q$ play a crucial role in Eq.~\ref{effhop}. This inter-band coupling cannot be neglected, as it directly influences the dynamics of the system. On the other hand, in the high-frequency limit, the Floquet bands decouple from each other, effectively reducing the system to the lowest-order Floquet band, which is captured by setting $p=q=0$ \cite{gomez-prl, delplace}. As a result, higher-order terms in $p$ and $q$ contribute minimally to Eq.~\ref{effhop}, and the coupling between the main lattice and its virtual copies becomes negligible.

In the low-frequency regime, the effective system size becomes quite large, making it challenging to achieve highly asymmetric transmission line-shapes. Consequently, our analysis focuses on the high-frequency limit, particularly under the condition $\hbar \omega \gg 4t$. In this regime, the light frequency is approximately $10^{16}\,$Hz, placing it in the near-ultraviolet to extreme ultraviolet range. The corresponding electric field strength is around $10^4\,$V/m, with a magnetic field of about $10^{-5}\,$T. The influence of the magnetic field generated by light is minimal, as its magnitude is extremely low. The light intensity used in this context is about $10^5\,$W/m$^2$, which is experimentally attainable. Notably, even higher light intensities have been used in recent studies, confirming that the selected intensity does not significantly alter the physical properties of the system~\cite{cwd1, cwd2}.

However, working within this high-frequency regime raises concerns about potential heating of the sample, a valid issue. This effect can be managed by utilizing artificial systems, as shown in previous research~\cite{delplace, polini}. These systems allow for the adjustment of lattice parameters, which could permit the use of lower frequencies and subsequently reduced field intensities, thereby mitigating heating concerns.

We now present our results, one by one, in different subsections. Before discussing the thermoelectric quantities, we first examine the structural behavior and band structure of the isolated BLGs (not attached to electrodes), to provide a clear understanding. 

\subsection{Structural properties and electronic band structure}
\begin{figure}[h]
\centering
\includegraphics[width=0.24\textwidth]{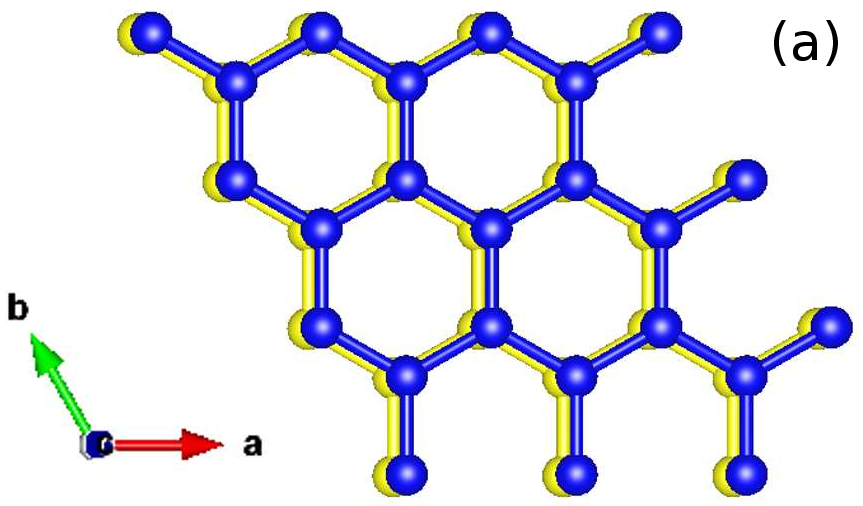}\hfill\includegraphics[width=0.24\textwidth]{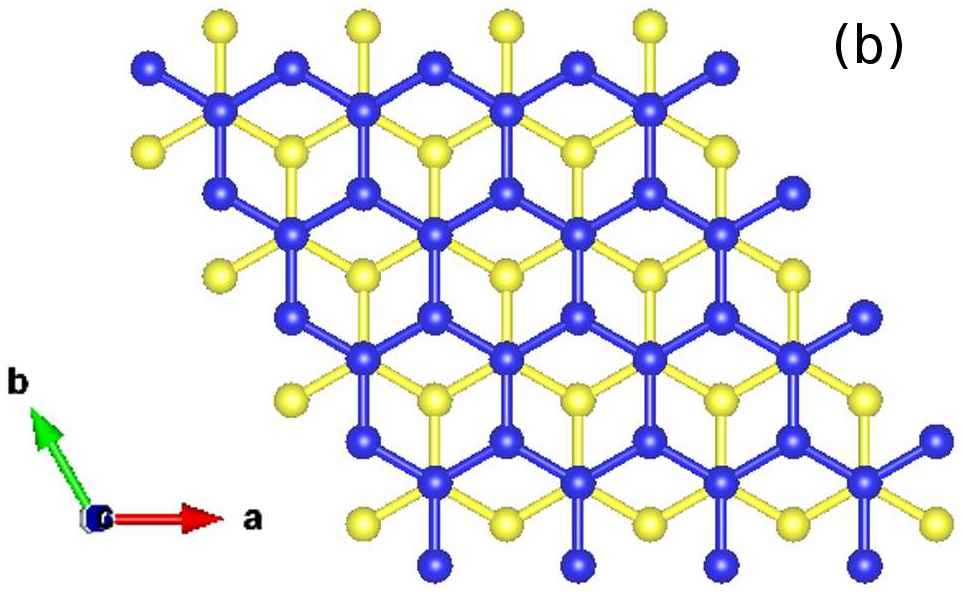}
\caption{(Color online). The geometric structures of (a) $AA$- and (b) $AB$-stacked BLGs. The blue and yellow balls denote the carbon atoms in the top and bottom layers, respectively.}
\label{schem}
\end{figure}

The geometric configurations of $AA$- and $AB$-stacked BLGs are depicted in Figs.~\ref{schem}(a) and (b), respectively. In $AA$-stacked BLG, the carbon atoms in both layers share identical $x$ and $y$ coordinates, forming an $AA$-stacked honeycomb lattice where all layers have the same configuration. This system exhibits $D_{6h}$ point group symmetry, which is characteristic of monolayer graphene~\cite{jiang-prb}. 
As illustrated in Fig.~\ref{schem}, the environment of carbon atoms in $AA$-stacked BLG differ significantly from those in $AB$-stacked system. In $AA$-stacked BLG, each carbon atom has three nearest-neighbors in each of the two adjacent layers. The optimized lattice parameters for $AA$- and $AB$-stacking BLGs are obtained to be  $a=b= 2.46\,$\AA, closely aligning with experimentally reported values ($a= 2.45\,$\AA)~\cite{razado}.

The $AB$-stacked bilayer BLG also forms a honeycomb lattice, where the bottom graphene layer is shifted relative to the top layer along one of their nearest C-C bonds in the horizontal plane, as depicted in Fig.~\ref{schem}(b). The $AB$-stacked BLG has lower symmetry than the $AA$-stacked BLG, belonging to the $D_{3d}$ point group~\cite{angeli}. The C-C bond length in both $AA$- and $AB$-stacked BLGs is 1.42$\,$\AA, which is nearly identical to the C-C bond length in bulk graphite~\cite{niyogi}. The interlayer distance $b$ differs slightly between the two configurations, measuring 3.49$\,$\AA~for $AA$-stacked BLG and 3.32$\,$\AA~for $AB$-stacked BLG. This variation in interlayer distance reflects differences in interlayer interactions due to the stacking configuration. Our calculations for the interlayer distance are consistent with experimental values, which are 3.55$\,$\AA~for $AA$-stacked graphene and 3.35$\,$\AA~for $AB$-stacked graphene~\cite{lee-jcp,hanfland}.

\begin{figure}[h]
\centering
\includegraphics[width=0.24\textwidth,height=0.21\textwidth]{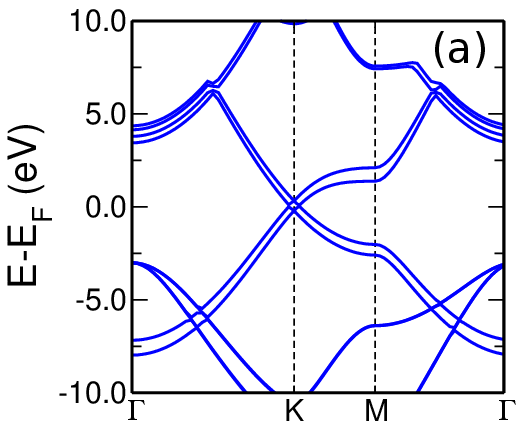}\hfill\includegraphics[width=0.24\textwidth,height=0.2\textwidth]{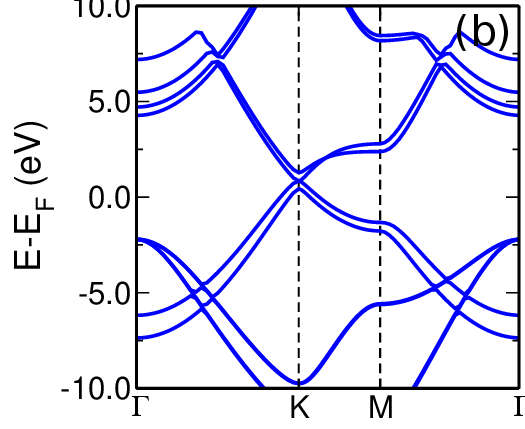}
\caption{(Color online). The energy dispersions are shown, respectively in (a) for the $AA$ stacking and (b) for the $AB$ stacking.}
\label{e-dis}
\end{figure}

The energy dispersions of $AA$- and $AB$-stacked BLGs are illustrated in Figs.~\ref{e-dis}(a) and (b), respectively. The electron distribution pattern of $AA$-BLG is six-fold~\cite{oudich}. It is very close to the free-standing monolayer graphene that produces a linear dispersion relation at $K$-point~\cite{yxu}. Whereas, the electron density function is a three-fold pattern in case of $AB$-stacked bilayer graphene. 
The difference in the related electron distribution function leads to the distinct energy band structure between the $AB$ and $AA$ bilayer graphene stacked layers. $AA$-stacked BLG exhibits linear and almost isotropic energy bands with slopes similar to monolayer graphene (MLG). The reason behind achieving liner dispersion in $AA$ stacking is the sublattice exchange having even symmetry and the interlayer electronic coupling suppressed by a significant Pauli repulsion arising between the graphene layers~\cite{yxu,Kindermann}. In contrast, $AB$-stacked BLG shows parabolic energy bands at the $K$-point due to interlayer coupling, leading to an overlap between the valence and conduction bands and disrupting their symmetry at the Fermi level~\cite{yxu}.

\subsection{Electronic transport properties}
Let us begin by discussing the electronic aspects of various thermoelectric quantities, starting with transmission probability, followed by electrical conductance, thermopower, and thermal conductance due to electrons. The nearest-neighbor hopping strength is considered as $t=2.7\,$eV and the coupling strength between the layers as $\gamma_0=0.4\,$eV for both type of stackings. The hopping parameters are chosen in alignment with those reported in the existing literature~\cite{hopref1,hopref2,hopref3,hopref4}.

\begin{figure}[h]
\centering
\includegraphics[width=0.5\textwidth]
{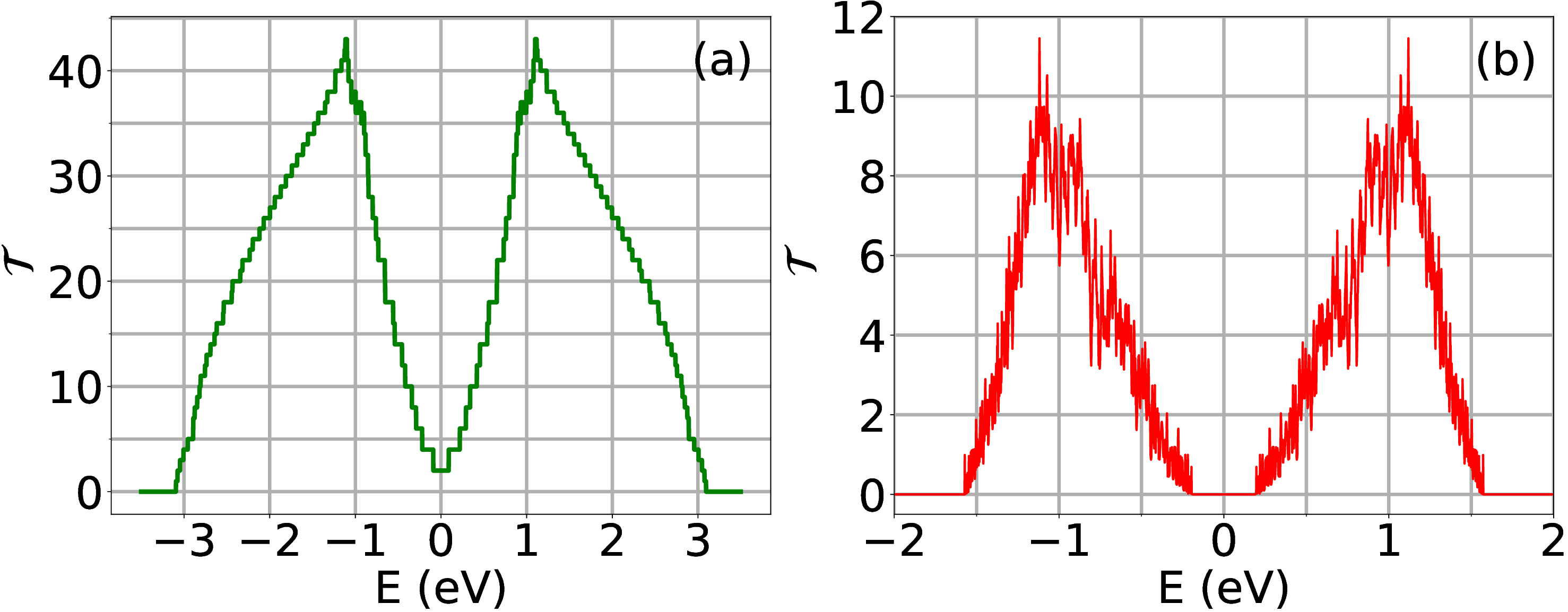} 
\caption{(Color online). Transmission probability ${\mathcal T}$ as a function of energy for $AA$-stacked BLG with zigzag edges. (a) without and (b) with irradiation. The light parameters are $A_x =1$, $A_y = 0.2$, and $\varphi=\pi/3$.}
\label{trans-aa}
\end{figure}

Figure~\ref{trans-aa} depicts the behavior of transmission probability as a function of energy for zigzag $AA$-stacked BLG. The length of the BLG is fixed at 5.7$\,$nm and width at 5.7$\,$nm. The widths of the electrodes are considered as same as that of the central system. 
In the absence of light, the transmission probability exhibits the typical step-like behavior (see Fig.~\ref{trans-aa}(a)), with each step corresponding to the opening of a new conduction channel as the energy increases. Near zero energy, the transmission probability remains finite, exhibiting metallic behavior owing to the zigzag configuration~\cite{waka} of the BLG considered in this study. However, when the system is irradiated with light, the transmission profile changes significantly as observed in Fig.~\ref{trans-aa}(b). The step-like behavior observed without light is completely lost, and the overall magnitude of the transmission probability decreases. This is due to the irradiation-induced spatial modulation through the renormalized hopping term, as described by Eq.~\ref{hop-light}. Additionally, the allowed energy window shrinks. In the absence of light, it spans from -3 to 3$\,$eV, but under irradiation, it reduces to approximately -1.5 to 1.5$\,$eV. Another noteworthy observation is the loss of metallic behavior around zero energy, which was present due to the zigzag edge in the absence of irradiation. Instead, a band gap forms in the presence of light. At the edges of this energy gap, the transmission probability sharply increases. This highly asymmetric transmission function is ideal for achieving a favorable thermoelectric response, as we will discuss later.

For the $AB$-stacked BLG, we also consider the identical system dimension as that of the $AA$-stacked BLG. A similar behavior is observed in $AB$-stacked BLG with zigzag edges. In the absence of light, 
\begin{figure}[h]
\centering
\includegraphics[width=0.5\textwidth]
{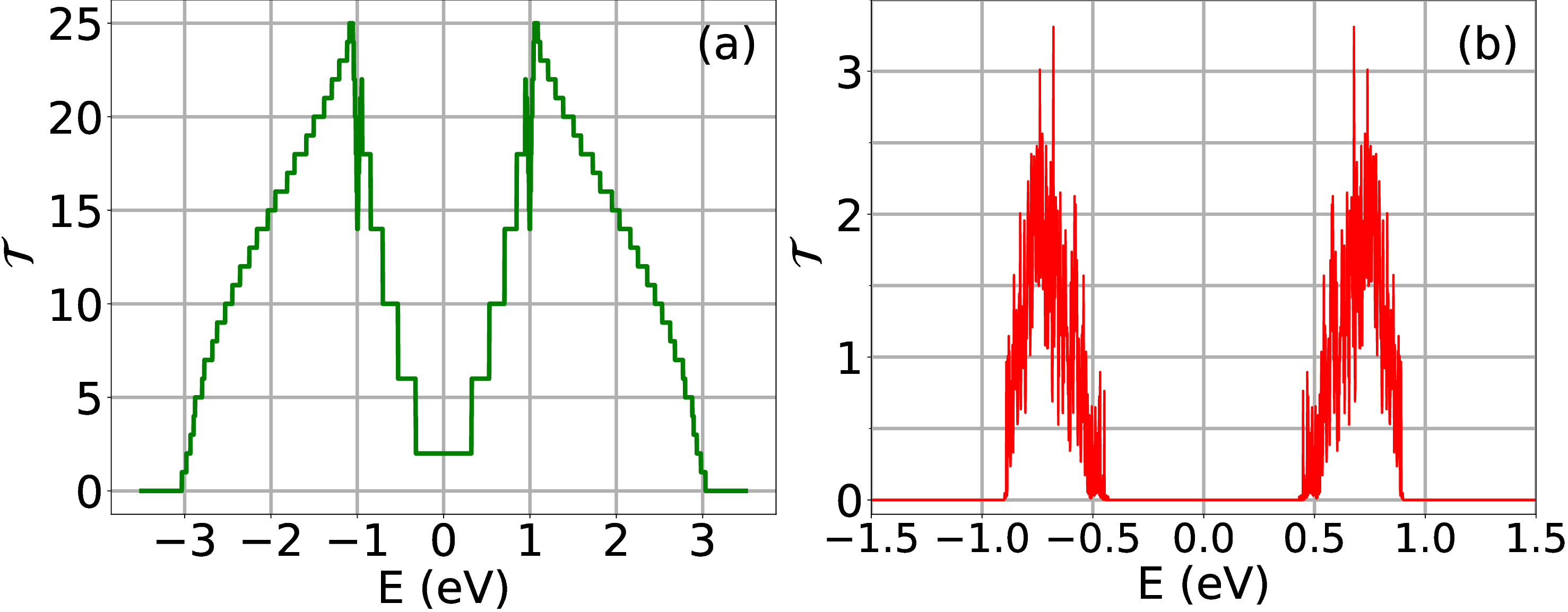} 
\caption{(Color online). Transmission probability ${\mathcal T}$ as a function of energy for $AB$-stacked BLG with zigzag edges. (a) without and (b) with irradiation. The light parameters are $A_x =1$, $A_y = 0.2$, and $\varphi=\pi/3$.}
\label{trans-ab}
\end{figure}
the transmission probability as a function of energy exhibits a step-like behavior, as shown in Fig.~\ref{trans-ab}(a). Compared to the $AA$-stacked BLG, the maximum transmission value is lower. This reduction is attributed to the lower symmetry of the $AB$ stacking, which permits fewer modes in the leads than in the $AA$-stacked BLG. When light is present, the step-like behavior is disrupted, and a band gap emerges, as illustrated in Fig.~\ref{trans-ab}(b). Consequently, $AB$-stacked BLG is also expected to exhibit a favorable TE response under irradiation.

\begin{figure}[h]
\centering
\includegraphics[width=0.5\textwidth]
{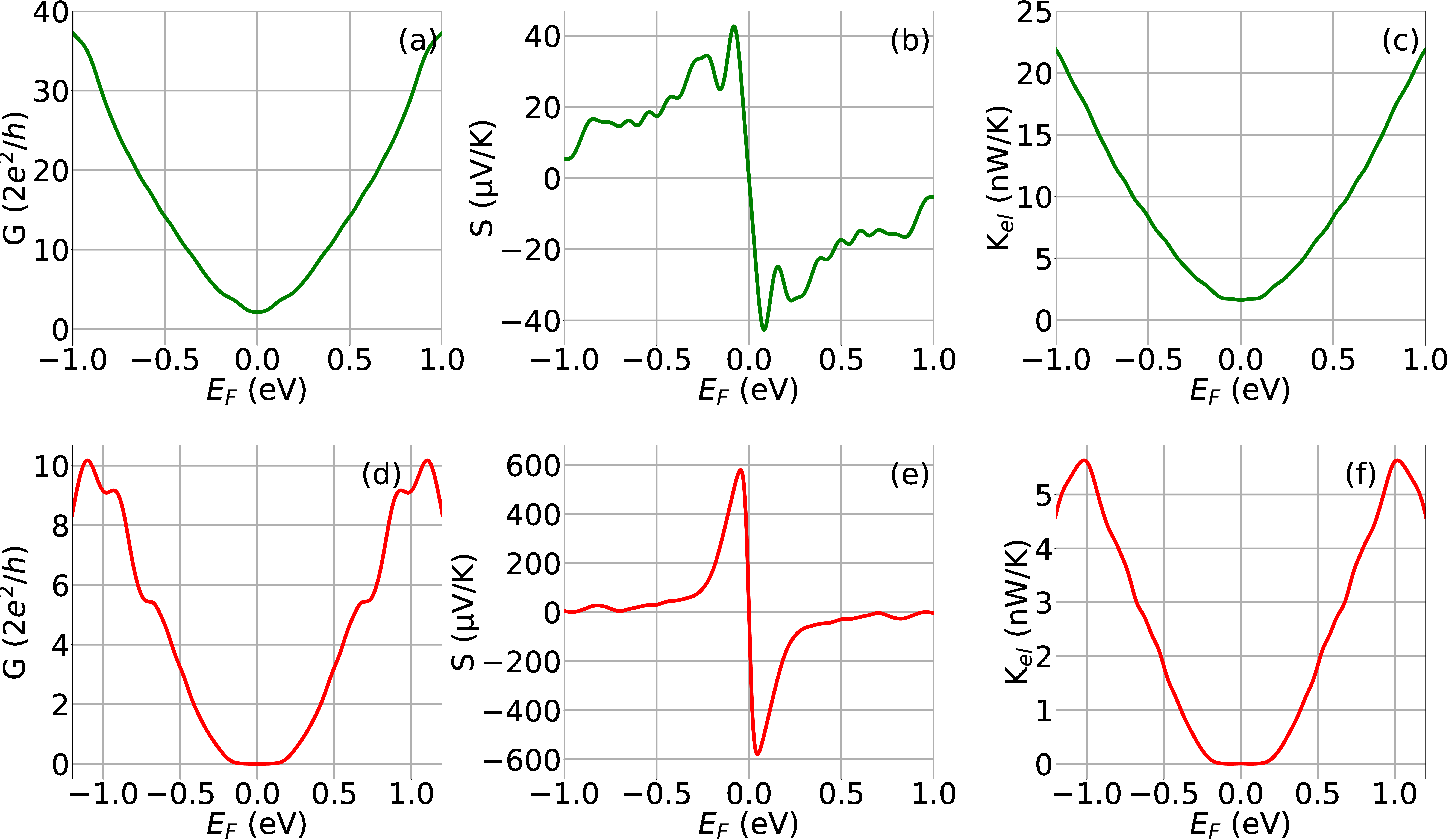} 
\caption{(Color online).  Different thermoelectric quantities as a function of Fermi energy for $AA$-stacked BLG. Top panel: in green (without irradiation). Bottom panel: in red (with irradiation). [(a), (d)] electrical conductance $G$. [(b), (d)] thermopower $S$. [(c), (f)] thermal conductance due to electrons $k_e$. The light parameters and all other system parameters are the same as in Fig.~\ref{trans-aa}. The temperature is fixed at $T=300\,$K.}
\label{gsk-aa}
\end{figure}
\begin{figure}[h]
\centering
\includegraphics[width=0.5\textwidth]
{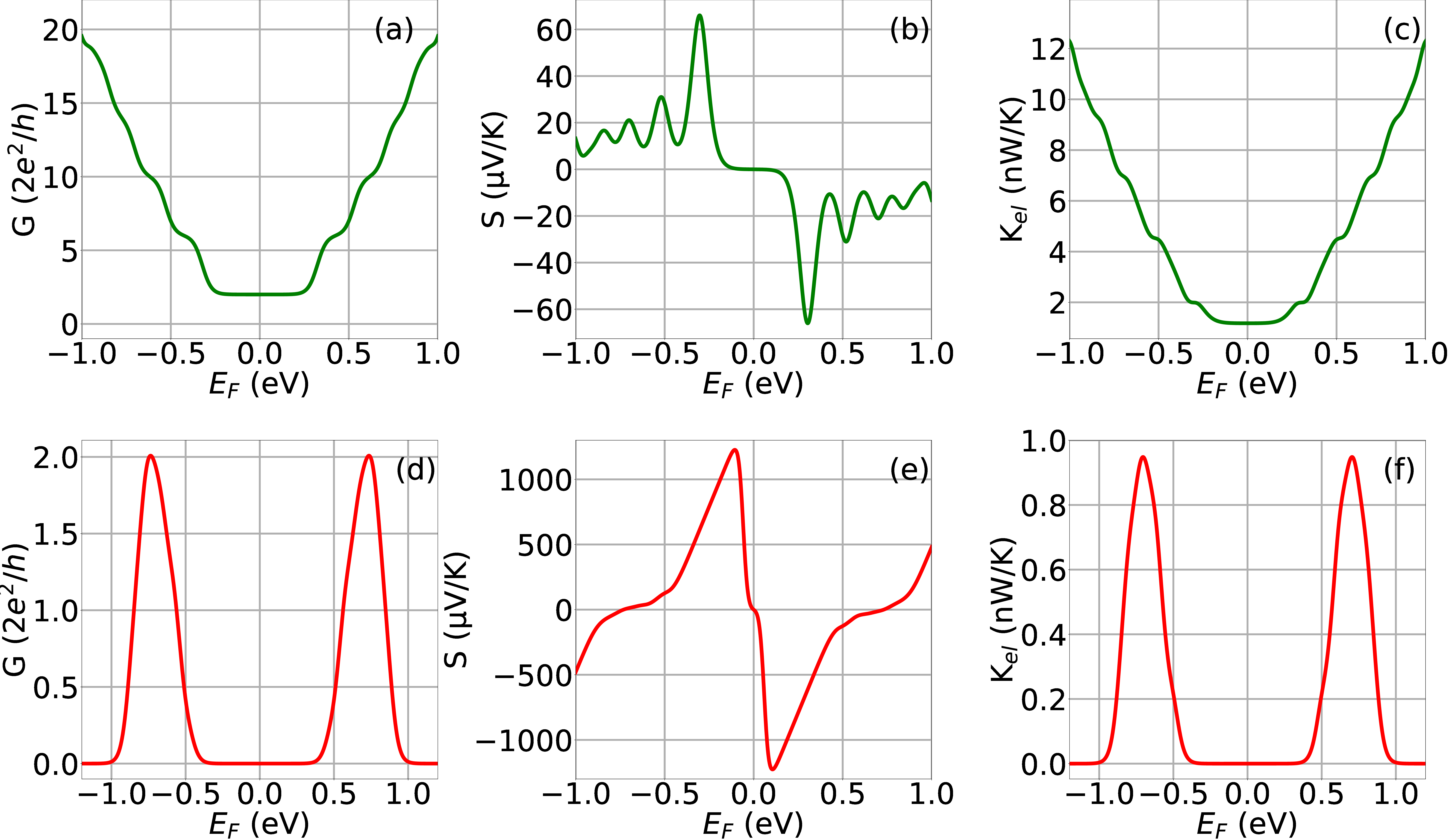} 
\caption{(Color online).  Different thermoelectric quantities as a function of Fermi energy for $AB$-stacked BLG. Top panel: in green (without irradiation). Bottom panel: in red (with irradiation). [(a), (d)] Electrical conductance $G$. [(b), (d)] thermopower $S$. [(c), (f)] thermal conductance due to electrons $k_e$. The light parameters and all other system parameters are the same as in Fig.~\ref{trans-aa}. The temperature is fixed at $T=300\,$K.}
\label{gsk-ab}
\end{figure}
With the computed transmission probability, we now study the thermoelectric quantities at room temperature $T = 300\,$K. We start by examining the behavior of the three TE quantities, namely, the electronic conductance $G$, thermopower $S$, and thermal conductance $K_e$ as a function of Fermi energy as depicted in Fig.~\ref{gsk-aa} and Fig.~\ref{gsk-ab} for the $AA$- and $AB$-stacked BLGs, respectively. The results without and with irradiation are depicted in the top and bottom panels, respectively. The light parameters and the system parameters are identical with Fig.~\ref{trans-aa}.

The behavior of electrical conductance, $G$ (in units of $2e^{2}/h$), as a function of Fermi energy $E_F$ is illustrated in Figs.~\ref{gsk-aa}(a) and (d), in the absence and presence of irradiation, respectively for the $AA$-stacked BLG. The conductance spectrum closely follows the $\mathcal{T}-E$ curve depicted in Fig.~\ref{trans-aa}, as $G$ is calculated using Eqs.~\ref{eq:10a} and \ref{eqn:11}. The conductance $G$ is plotted within a narrow energy range of -1 to 1. In the absence of light, the sharp, step-like features are smoothed out due to thermal broadening. When light is present, the magnitude of $G$ is significantly reduced, by approximately a factor of 4, compared to the case without light. In the presence of light, near the zero Fermi energy, conductance becomes vanishingly small. For $AB$-stacked BLG, the maximum conductance in the absence of light (Fig.~\ref{gsk-ab}(a)) is noticeably lower than that of $AA$-stacked BLG, as evident from their respective transmission spectra. As shown in Fig.~\ref{gsk-ab}(d), the electrical conductance decreases by nearly 10-fold in the presence of irradiation.

The thermopower $S$ is calculated using Eq.~\ref{eq:10b}, with its corresponding thermal integral, $L_1$, derived from Eq.~\ref{eqn:11}. In the absence of irradiation, for $AA$-stacked BLG, $S$ (as depicted in Fig.~\ref{gsk-aa}(b)) is approximately $40\,\mu$V/K, which is insufficient for effective thermoelectric applications. However, under irradiation, a substantial increase in $S$ is observed in the $AA$-stacked BLG (as shown in Fig.~\ref{gsk-aa}(e)), reaching arount $600\,\mu$V/K, nearly ten times greater than that for the non-irradiated case. This significant enhancement in $S$ can greatly improve the FOM, as $ZT \propto S^2$. The favorable response is attributed to the irradiation effect, which alters the intralayer hopping integrals and introduces greater asymmetry in the transmission spectrum around the edge of the bandgap. For $AB$-stacked BLG, the Seebeck coefficient ($S$) is relatively low in the absence of light, as shown in Fig.~\ref{gsk-ab}(b), with a maximum magnitude of approximately $60\,\mu$V/K. However, in the presence of light (Fig.~\ref{gsk-ab}(e)), $S$ increases dramatically, more than tenfold, reaching a maximum value of around $1000 \,\mu$V/K. This value is significantly higher than that observed in irradiated $AA$-stacked BLG. In all the cases, $S$ is antisymmetric about $E_F=0$ due to the electron-hole symmetry inherent in the system.

The thermal conductance due to electrons $k_e$ exhibits a behavior similar to that of the electronic conductance as a function of Fermi energy. Notably, $k_e$ reaches lower values under irradiation (Fig.~\ref{gsk-aa}(f)) compared to the irradiation-free case (Fig.~\ref{gsk-aa}(c)) for the $AA$-stacked BLG. For the $AB$-stacked BLG, we also observe the similar behavior both in the absence and presence of light as shown in Figs.~\ref{gsk-ab}(c) and (f), respectively. This reduction in $k_e$ due to light irradiation is beneficial for thermoelectric performance, as a lower $k_e$ directly contributes to an increase in $ZT$, where $k_e$ is in the denominator of the expression.

\subsection{Phonon transport properties}
Before examining the phonon transport properties, it is essential to highlight that the irradiation frequency is approximately $10^{15}\,$Hz, whereas the phonon vibration frequency is around $10^{12}\,$Hz. This indicates that the irradiation frequency is roughly three orders of magnitude higher than the phonon vibrational frequency. As a result, we can reasonably assume that irradiation does not significantly affect lattice vibrations. Therefore, the impact of irradiation on lattice vibrations can be safely neglected in our analysis.

\begin{figure}[h]
\centering
\includegraphics[width=0.24\textwidth]{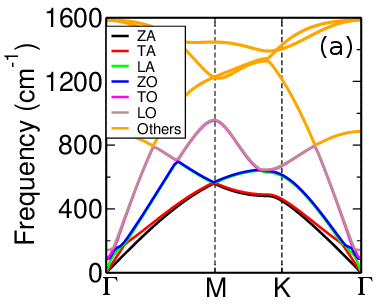}\hfill\includegraphics[width=0.24\textwidth]{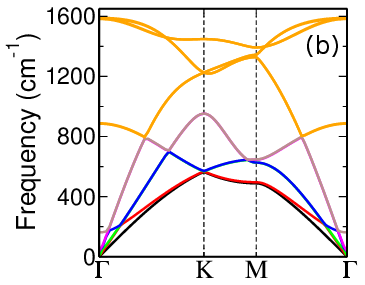} 
\caption{(Color online). Phonon dispersions of (a) AA-BLG and (b) AB-BLG along high-symmetry $q$-points in the hexagonal Brillouin zone. }
\label{ph-dis}
\end{figure}

A comparative analysis of phonon dispersions along the high-symmetric points in the
irreducible hexagonal Brillouin zone (BZ) for $AA$- and $AB$-stacked BLGs is
shown in Figs.~\ref{ph-dis}(a) and (b), respectively, exhibiting excellent agreement with existing experimental and theoretical data~\cite{kong,coce}. Both stacking configurations share similar mode characteristics, with notable differences at the $\Gamma$ point. For both systems, transverse acoustic
(TA) and longitudinal acoustic (LA) modes display linear dispersion at low $q$-vectors, whereas
the ZA mode, showing quadratic dispersion near the $\Gamma$ point with out-of-plane atomic
displacements. This quadratic dependence can be attributed to the two-dimensional out-of-plane phonon mode and three-fold rotational symmetry for $AB$-BLG and six-fold for $AA$-BLG~\cite{dsuza}. Our observations confirm that not only the LA and TA mode frequencies near the BZ centre are lower in $AB$-BLG compared to $AA$-BLG, further the corresponding layer breathing mode (ZO) found at 84$\,$cm$^{-1}$ and 36$\,$cm$^{-1}$ for $AA$-BLG and $AB$-BLG, respectively, aligning with already existing
reported data~\cite{kong}. It is further important to note that LA and ZO remain indistinguishable
throughout the BZ, which could cause more scattering rate among them for both $AA$- and $AB$-BLGs, which will be discussed later. Similar to previous studies, we also report that the acoustic and
layer breathing modes exhibit linear and parabolic dispersions in $AA$- and $AB$-BLGs,
respectively, near the $K$-point. It is noteworthy here to highlight that the phonon spectra near
the $K$-point resembles similarities to the electron spectra near $K$-point in $AA$- and $AB$-BLG as
shown in Figs.~\ref{e-dis}(a) and (b).

The lattice thermal conductivity $\kappa_{\text {latt}}$ is calculated by summing the contributions of all phonon modes denoted by the wave vector $q$ using ShengBTE code~\cite{bte}, which solves phonon Boltzmann transport equation (PBTE) given as
\begin{equation}
\kappa_{\text {latt}}^{\alpha\beta} = \frac{1}{k_B T^2 \Omega N}\sum_\lambda f_0 \left(f_0 + 1\right) \left(\hbar\omega_\lambda\right)^2 v_\lambda^\alpha F_\lambda^\beta,
\label{klat}
\end{equation}
where, $N$, $\Omega$, $f_0$, and $v_\lambda$ are the number of uniformly spaced $q$ points in the Brillouin zone, volume of the unit cell, the Bose-Einstein distribution function depends upon phonon frequency $\omega_\lambda$  and the phonon group velocity, respectively. $F_\lambda^\beta = \tau_\lambda^\beta v_\lambda$ where, $\tau_\lambda^\beta$ is the phonon relaxation time. 

\begin{figure}[h]
\centering
\includegraphics[width=0.24\textwidth,height=0.21\textwidth]{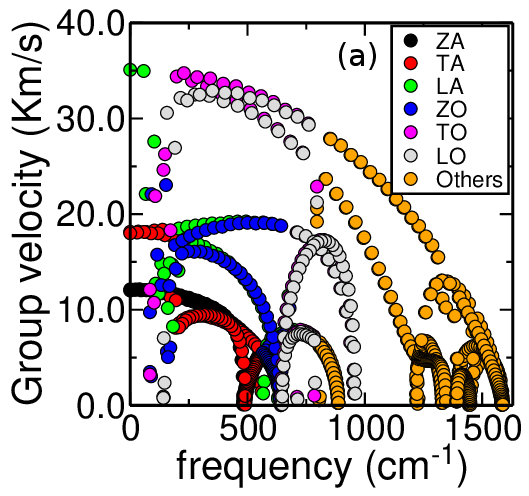}\hfill\includegraphics[width=0.24\textwidth,height=0.21\textwidth]{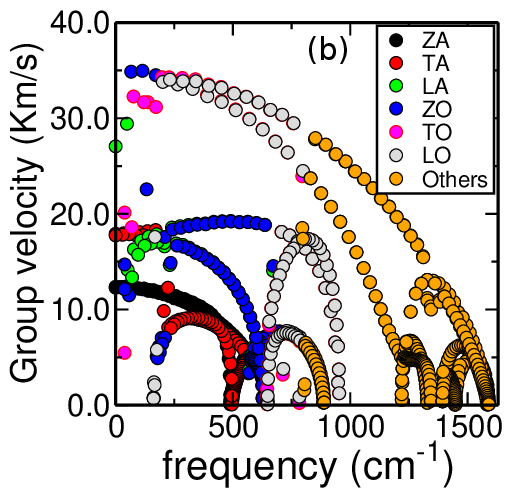}\vskip 0.05 in
\includegraphics[width=0.24\textwidth,height=0.21\textwidth]{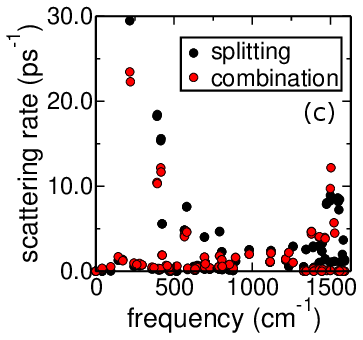}\hfill\includegraphics[width=0.24\textwidth,height=0.21\textwidth]{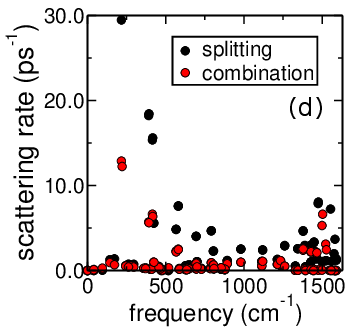}\vskip 0.05 in
\includegraphics[width=0.24\textwidth,height=0.21\textwidth]{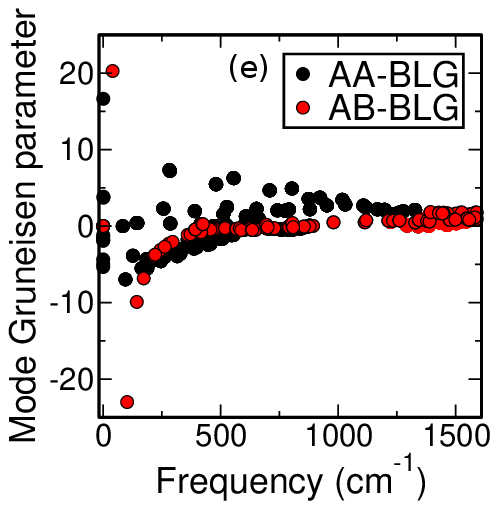}\hfill\includegraphics[width=0.24\textwidth,height=0.21\textwidth]{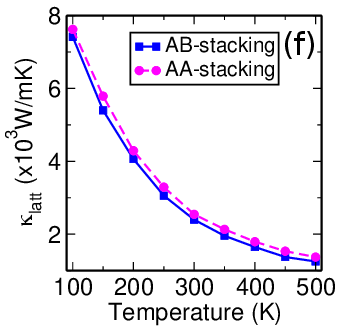}
\caption{(Color online). Calculated group velocity and scattering rate of acoustic (ZA, TA, LA) and low frequency optical (ZO, TO, LO) phonons and other optical phonons for (a), (c) $AA$- and (b), (d) $AB$-BLG, respectively, as a function of frequency. (e) The comparison of mode Gr\"{u}neisen parameter as a function of frequency for $AA$- and $AB$-BLG and (f) the lattice thermal conductivity $\kappa_{\text {latt}}$ of $AA$- and $AB$-stacked BLG as a function of temperature.}
\label{vg}
\end{figure}

To understand the role of each vibrational mode in the observed lattice thermal conductivity, we calculate the mode Gr\"{u}neisen parameter, group velocity, and scattering rate. The phonon group velocity is determined from the phonon spectrum, expressed as,
\begin{equation}
v_{\lambda,q} = \frac{\partial\omega_{\lambda,q}}{\partial q},
\end{equation}
where $\omega_{\lambda,q}$ denotes the phonon frequency. The phonon group velocity for $AA$- and $AB$-stacked BLGs is calculated and shown in Figs.~\ref{vg}(a) and (b), respectively, which is consistent with other theoretical results~\cite{duan}. Both systems exhibit high lattice thermal conductivity primarily due to its elevated phonon group velocities. Notably, the maximal phonon group velocity of the longitudinal acoustic (LA) branch reaches up to 35$\,$km/s for $AA$-BLGs, surpassing most of other two-dimensional materials such as silicene and germanene~\cite{peng}. The longitudinal acoustic (LA) and transverse acoustic (TA) phonon modes have higher group velocities compared to the out-of-plane acoustic (ZA) mode, due to its quadratic dispersion behaviour in both the materials. The highest group velocities of the LA, TA and ZA modes in $AA$-BLG ($AB$-BLG) correspond to 35.0 (30.0), 18.0 (17.8) and 12.7 (11.0) km/s, respectively. Interestingly, the $AB$-BLG exhibits lower group velocities than $AA$-BLG, which is likely to lead lower lattice thermal conductivity, as acoustic modes play significant role in determining $\kappa_{\text {latt}}$. Moreover, it is important to highlight that the acoustic mode group velocities in $AB$-BLG are even lower than those of the optical modes (ZO, TO, and LO), suggesting further reduction in $\kappa_{\text {latt}}$ for $AB$-BLG as mentioned in Fig.~\ref{vg}(b).

To accurately compute $\kappa_{\text {latt}}$, effects from the harmonic and anharmonic lattice displacements
should be considered to include contributions of second and third order phonon-phonon scattering processes, respectively. As the Gr\"{u}neisen parameter ($\gamma$) provides anharmonic interactions strength between lattice vibrations and the degree of phonon scattering, therefore we calculate the mode-dependent Gr\"{u}neisen parameter ($\gamma$) for $AA$- and $AB$-BLG, which
characterizes the relationship between phonon frequency and crystal volume change. The Gr\"{u}neisen parameter $\gamma$ is expressed as
\begin{equation}
\gamma_\lambda = \frac{-V}{\omega_\lambda}\frac{\partial\omega_\lambda}{\partial V},
\end{equation}
where $\omega_\lambda$ and $V$  denotes the phonon frequency and crystal volume, respectively. It is important to
mention the positive $\gamma$ corresponds to a decrease in the phonon frequency with increasing
volume and the inverse is true for negative $\gamma$. The frequency-dependent mode Gr\"{u}neisen
parameters for both the systems presented in the Fig.~\ref{vg}(e) reveal the large phonon scattering
strength with $\gamma\sim 20$ observed in the $AB$-stacked BLG is attributed to the presence of strong
anharmonicity within the layered structure. In the low frequency regime especially below 500$\,$cm$^{-1}$ , the mode resolved $\gamma$ for the acoustic modes and the low lying optical modes (ZO, TO,
LO) considerably surpass than those of the high frequency optical modes, shown in Figs.~\ref{s1}(a) and (b) of supporting material, which is consistent with previous studies~\cite{kong,duan}. This
implies acoustic modes and low energetic optical modes are heavily dispersed with higher
scattering strength, leading to higher anharmonicity for these specific modes. Additionally, it
is interesting to note here that the optical phonon modes lying between 0-150$\,$cm$^{-1}$ range show
high Gr\"{u}neisen parameters exceeding 20, suggesting high anharmonicity of certain low-lying
optical modes in $AB$-BLG. The high-frequency optical phonon modes possess relatively lower Gr\"{u}neisen parameters ranging from -0.68 to +2.5. The relatively higher $\gamma$ of $AB$-BLG could be
a contributing factor in reducing lattice thermal conductivity of $AB$-BLG relative to $AA$-BLG.

Furthermore, it is evident from Figs.~\ref{vg}(c) and (d) that the higher scattering rates below 500$\,$cm$^{-1}$ occurring between the acoustic and low-frequency optical phonons signify enhanced
anharmonic process and responsible for reducing phonon relaxation time in both the structures.
The combination process of three phonon scattering rates dominates at low frequency (below
800$\,$cm$^{-1}$), and gradually coexists with the splitting process between 500 to 1580$\,$cm$^{-1}$, while
the splitting process abounds above 1250$\,$cm$^{-1}$. There is a scattering peak around 250 to 500$\,$cm$^{-1}$ 
because of the presence of indistinguishable layer breathing mode and LA branch mentioned
in Figs.~\ref{ph-dis}(a) and (b), thereby enhancing phonon-phonon scattering rate.

Finally, we obtain the $\kappa_{\text {latt}}$ for temperatures ranging from 100 to 500$\,$K for $AA$- and $AB$-stacked BLGs using Eq.~\ref{klat} as implemented in the ShengBTE code~\cite{bte} and shown in Fig.~\ref{vg}(f) and
Table~\ref{table} of supporting material. The gradually decrease in $\kappa_{\text {latt}}$ with increasing temperature is
attributed to heightened anharmonic phonon scattering rate and higher anharmonic scattering
strength consistent with the anticipated inverse relationship between $\kappa_{\text {latt}}$ and temperature. The calculated $\kappa_{\text {latt}}$ values for $AA$- and $AB$-stacked BLGs are 2444$\,$W/mK and 2396$\,$W/mK, respectively, at
300$\,$K (align with experimental results of 1896.2 $\pm$ 410$\,$W/mK~\cite{hli}). The characteristic behaviour
of $AA$- and $AB$-stacked BLGs are notably similar across the temperature
range of 100-500$\,$K, both displaying nearly identical trend in lattice thermal conductivities.
The predominance of lower group velocity and strong anharmonic scattering strength of
acoustic and low-lying optical modes are responsible in reducing lattice thermal conductivity
in $AB$-BLG compared to $AA$-BLG.

\subsection{Figure of merit}
With the computed electronic thermoelectric quantities derived from the tight-binding model and the phonon thermal conductivity from DFT analysis, we are now ready to calculate the FOM in the presence of light. 
\begin{figure}[h]
\centering
\includegraphics[width=0.48\textwidth]{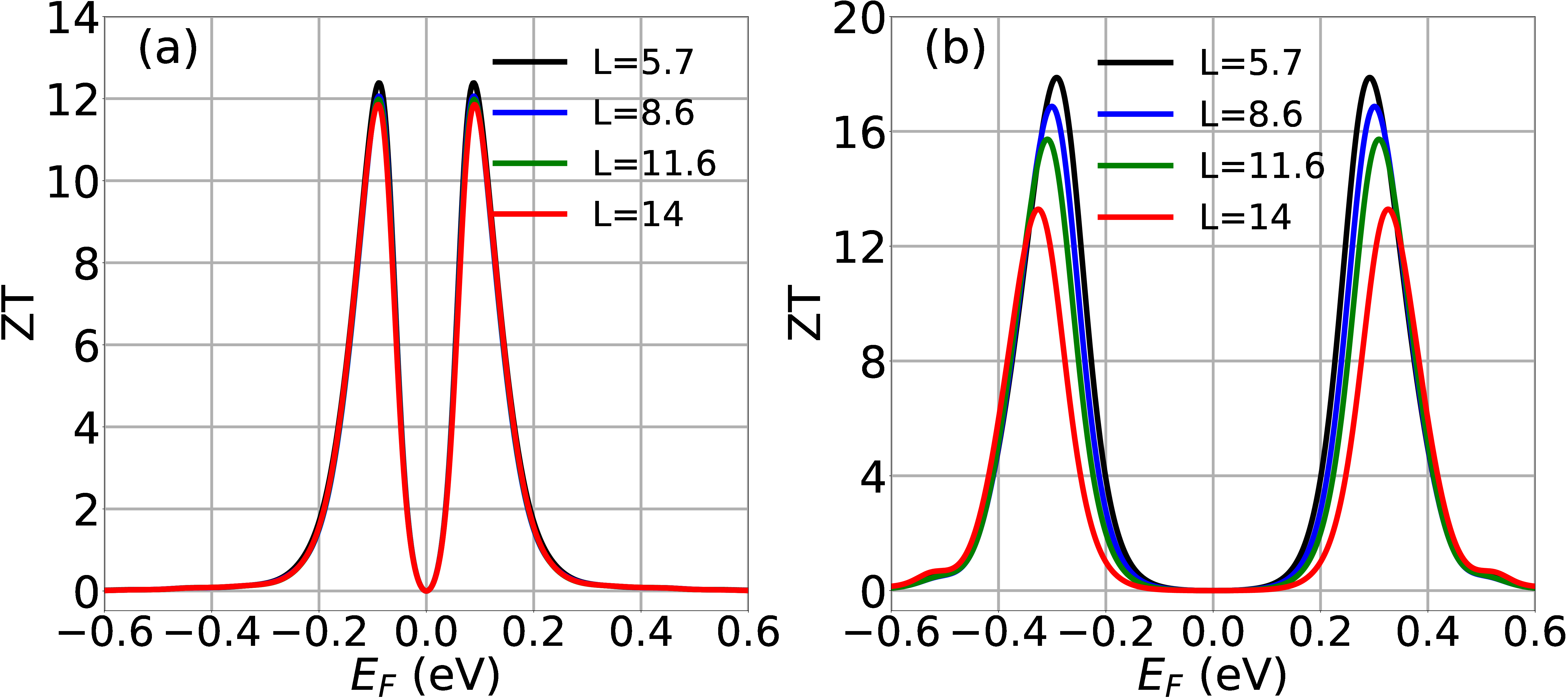}
\caption{(Color online). $ZT$ as a function of the Fermi energy $E_F$ for (a) $AA$- and (b) $AB$-stacked BLG. Four different lengths are considered, namely 5.7, 8.6, 11.6, and 14$\,$nm, with corresponding results represented by black, blue, green, and red colors. The width is kept constant for all the cases and is 5.7$\,$nm. All other system parameters and the light parameters are the same as in Fig.~\ref{trans-aa}.}
\label{fom}
\end{figure}
In Figs.~\ref{fom}(a) and (b), we plot $ZT$, as a function of Fermi energy $E_F$ for $AA$- and $AB$-stacked BLGs, respectively. We consider four different lengths of BLG, namely, 5.7, 8.6, 11.6, and 14$\,$nm, while keeping the width constant at 5.7$\,$nm for all lengths. The computed phonon thermal conductances for these lengths are 0.078, 0.12, 0.16, and 0.19$\,$pW/K, respectively, with corresponding results represented by black, blue, green, and red colors. All other physical and light parameters remain constant with those mentioned in Fig.~\ref{trans-aa}.

As the system length increases, the change in $ZT$ for $AA$-stacked BLG is minimal. However, for $AB$-stacked BLG, $ZT$ systematically decreases with increasing length. This is expected, as the phonon thermal conductivity, $k_{\text{ph}}$, increases with length, leading to a reduction in $ZT$. Interestingly, the maximum $ZT$ is found to be higher for $AB$-stacked BLG compared to $AA$-stacked BLG. Overall, under the considered set of light parameters, both types of BLGs exhibit a favorable thermoelectric response, with $ZT$ values exceeding much higher than unity.

\begin{figure}
\centering
\includegraphics[width=0.48\textwidth]{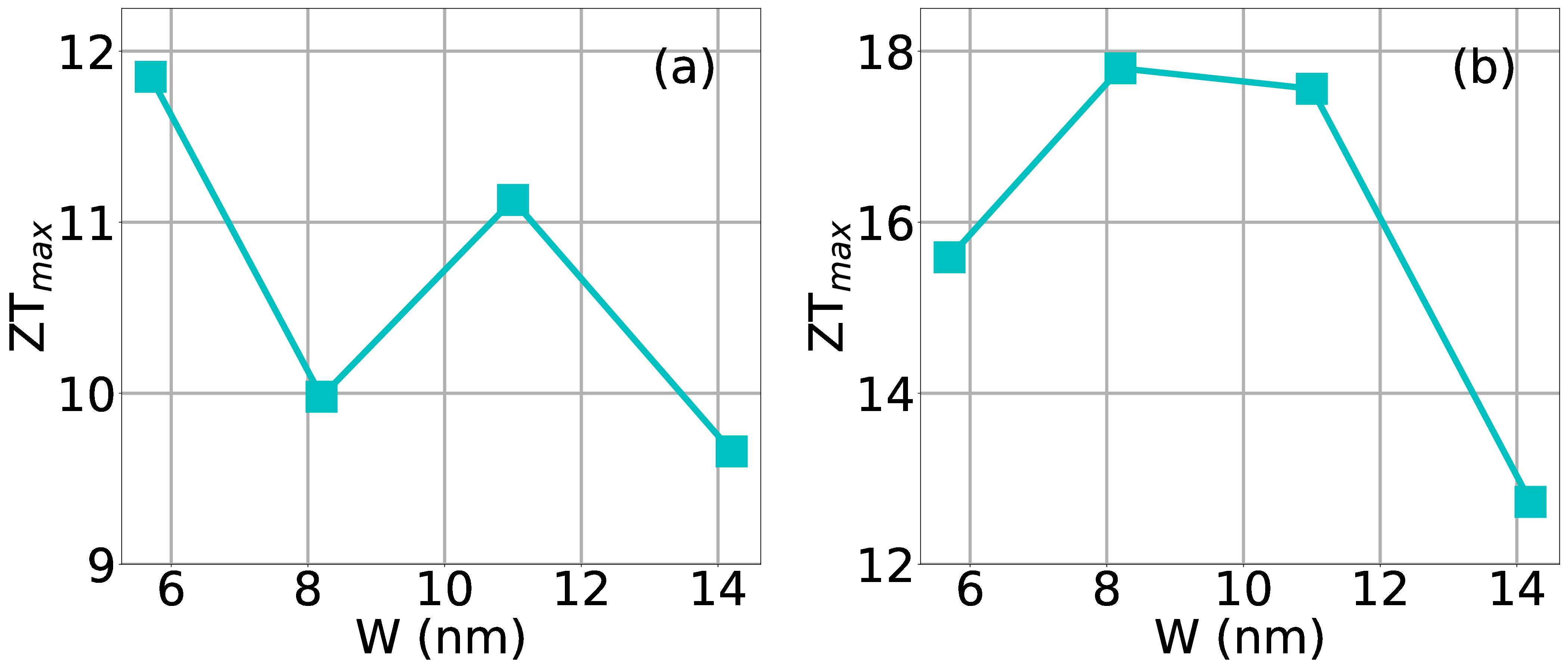}   
\caption{(Color online). $ZT_{max}$ as a function of width $W$ for (a) $AA$- and (b) $AB$- stacking. The length is kept constant for all the cases and is 14$\,$nm. All other system parameters and the light parameters are the same as in Fig.~\ref{trans-aa}.}
\label{wvar}
\end{figure}
Next, we examine how varying the width of the BLG affects the behavior of $ZT$. Specifically, we compute the maximum $ZT$, denoted as $ZT_{max}$, over the Fermi energy window from -1 to 1$\,$eV. 
The results are presented in Figs.~\ref{wvar}(a) and (b) for the $AA$- and $AB$-stacked BLGs, respectively, with a fixed length of 14$\,$nm for both types of BLG. Remarkably, in the presence of irradiation, $ZT$ remains consistently high across all considered widths. For the $AA$-stacked configuration, $ZT$ is always greater than 9, while for the $AB$-stacked configuration, it exceeds 12. This demonstrates that high $ZT$ values can be achieved across various BLG widths when irradiation is applied.

\begin{figure}[h]
\centering
\includegraphics[width=0.5\textwidth]{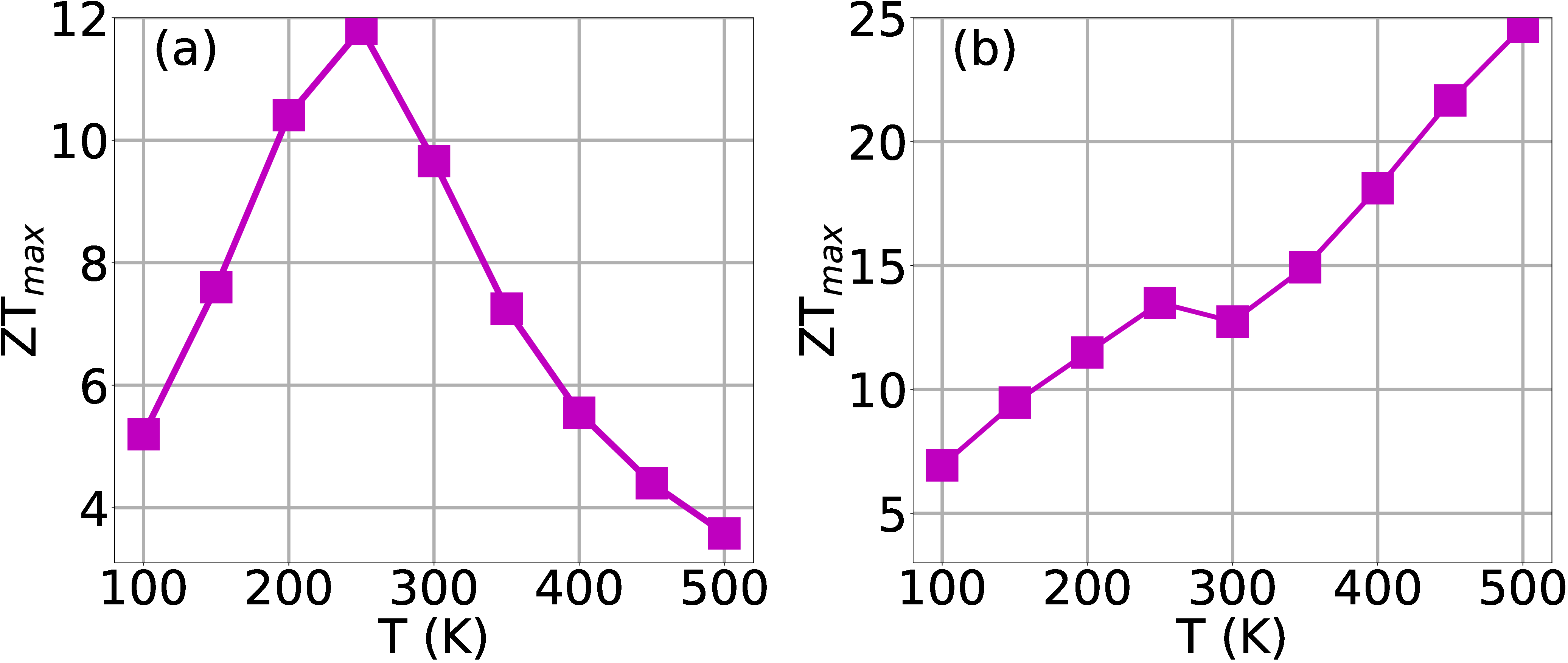}   
\caption{(Color online). $ZT_{max}$ as a function of temperature $T$ for (a) $AA$- and (b) $AB$- stacking. All the system parameters and the light parameters are the same as in Fig.~\ref{trans-aa}.}
\label{tempvar}
\end{figure}
All the TE quantities thus far have been calculated at room temperature, $T = 300\,$K. To explore how $ZT$ varies with temperature, we plot $ZT_{max}$ as a 
function of temperature $T$ for both $AA$- and $AB$-stacked BLGs, as shown in Figs.~\ref{tempvar}(a) and (b), respectively. The definition of $ZT_{max}$ remains the same as before. In this analysis, the temperature is varied from 100 to 500$\,$K.
For the $AA$-stacked BLG, $ZT$ initially increases and show a decreasing trend beyond $T \sim 250\,$K. In contrast, $ZT$ in the $AB$-stacked BLG exhibits a consistent upward trend with increasing temperature. A key observation is that $ZT$ remains greater than unity across the entire temperature range, which is promising from an application standpoint.  

We also observe favorable thermoelectric (TE) responses under different light polarization parameters. Specifically, we considered three sets of light parameters: $A_x = 1$, $A_y = 0.3$, $\varphi = \pi$; $A_x = 0.5$, $A_y = 0.2$, $\varphi = \pi/1$; and $A_x = 1$, $A_y = 0.3$, $\varphi = \pi/6$, and computed $ZT$ for both $AA$- and $AB$-stacked BLGs. The results, presented in Fig.~\ref{s2} of the supporting material, show that $ZT$ exceeds unity at room temperature.

Before concluding, it is important to highlight that all the results discussed for the BLGs pertain to systems with zigzag edges. A common question is whether the same approach applies to armchair edges, which can exhibit conditional metallicity--depending on their width, they may be metallic or semiconducting with a finite band gap~\cite{waka}. In this study, we demonstrated that irradiation can induce a gap in zigzag-edged BLGs, and this effect is equally applicable to semiconducting armchair edges. Similarly, for metallic armchair edges, irradiation can also induce a gap. Therefore, it is quite expected that irradiation will have a beneficial effect in achieving a favorable thermoelectric response in BLGs with armchair edges, irrespective of their metallicity.
 
\section{\label{sec4}Summary}
In summary, we have proposed a method to achieve a high figure of merit in bilayer graphene flakes by irradiating them with arbitrarily polarized light. We have calculated various thermoelectric quantities, including electrical conductance, thermopower, and thermal conductance due to electrons, using a tight-binding theory that incorporates the effects of irradiation. Additionally, we computed the thermal conductance due to phonons using density functional theory. Combining these results, we have determined the FOM, which exhibits a highly favorable response, consistently exceeding unity. The BLGs are considered for two types, namely $AA$- and $AB$-stacked with zigzag edges.

For the electronic part, we have employed KWANT to get the electronic transmission probability. We observed that the transmission probability is significantly altered by irradiation, leading to the formation of a gap around zero energy in both types of BLG flakes. As a result, the electrical conductance and the thermal conductance due to electrons decrease around the zero Fermi energy. Conversely, the Seebeck coefficient increases substantially, whereas it was notably low in the absence of light. The thermal conductance due to electrons near the gap is only about a few pW/K for both type of stackings, while the thermopower is about 600$\,\mu$V/K for $AA$-stacked and about 1000$\,\mu$V/K for $AB$-stacked BLG.

Using the DFT method combined with Boltzmann transport theory, we have calculated the
structural and electronic properties of $AA$- and $AB$-BLGs. $AA$-stacked BLG exhibiting linear
energy bands, while $AB$-stacked BLG shows parabolic energy bands at the $K$-point. We
have investigated phonon dispersion of both stacking configurations using DFPT and obtained thermal
transport coefficient employing {\it ab-inito} theory of the anharmonic properties, that includes
third-order phonon-phonon scattering. Our study systematically showed that the lattice thermal
conductivity of $AB$-stacked BLG is slightly lower compare to $AA$-stacked, attributed to
enhancement of anharmonic scattering strength and surpassing phonon group velocity
observed in $AB$ configuration. 

Combining the results for the electronic and phononic parts, we finally have computed $ZT$ in the presence of irradiation, which exhibits highly favorable response, exceeding unity in both kinds of stackings. We further studied the behavior of $ZT$ with varying length and width of the system and achieved higher FOM. A temperature variation has also been discussed and we got very good response over the considered temperature range. Notably, we observed an enhanced FOM under different light polarizations, which significantly strengthens the practical applicability of this work.

\setcounter{secnumdepth}{0}
\section{ACKNOWLEDGMENT}

RKB thankfully acknowledges the financial support of the Science and Engineering Research Board, Department of Science and Technology, Government of India (Project File Number: EEQ/2022/000325).

\pagebreak
\widetext
\begin{center}
\textbf{\large Supplemental Materials: Thermolectricity in irradiated bilayer graphene flakes}
\end{center}
\setcounter{equation}{0}
\setcounter{figure}{0}
\setcounter{table}{0}
\setcounter{page}{1}
\makeatletter
\renewcommand{\theequation}{S\arabic{equation}}
\renewcommand{\thefigure}{S\arabic{figure}}
\renewcommand{\bibnumfmt}[1]{[S#1]}
\renewcommand{\citenumfont}[1]{S#1}
\begin{figure}[h]
\centering
\includegraphics[width=0.4\textwidth]{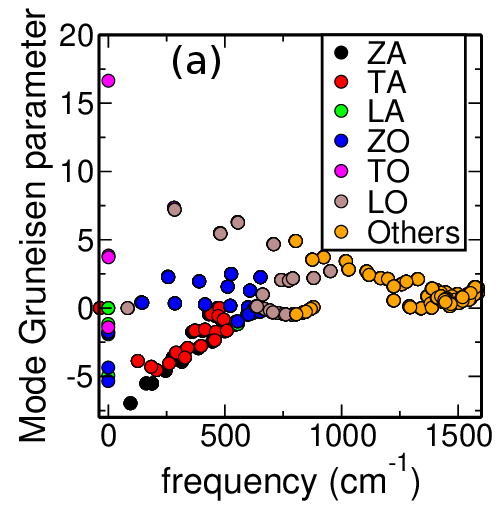}\hfill\includegraphics[width=0.4\textwidth]{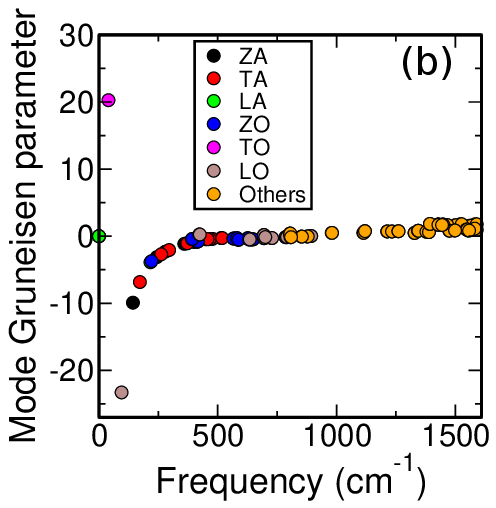}
\caption{(Color online). Calculated mode Gr\"{u}neisen parameter for (a) $AA$-BLG and (b) $AB$-BLG.}
\label{s1}
\end{figure}

\begin{table}[h!]
\caption{\label{table}The calculated lattice thermal conductivity for $AA$- and $AB$-stacked BLGs for
temperature range of 100 to 500$\,$K.}
\begin{center}
\begin{tabular}{ |c|c| c| }
\hline
\hline
 Temperature & $AA$-BLG & $AB$-BLG\\
      (K)          & $\kappa_{\text {latt}}$ (W/mK)     & $\kappa_{\text {latt}}$ (W/mK)       \\
\hline
\hline
 100 & 7622 & 7414\\
 \hline
150 & 5788 & 5398\\
 \hline
200 & 4011 & 4063\\
 \hline
 250 & 3072 & 3057\\
 \hline
 300 & 2444 & 2396\\
 \hline
 350 & 2032 & 1954\\
 \hline
 400 & 1743 & 1646\\
 \hline
 450 & 1532 & 1372\\
 \hline
 500 & 1371 & 1250\\
 \hline
\end{tabular}
\label{table}
\end{center}
\end{table}

\begin{figure}[h]
\centering
\includegraphics[width=0.6\textwidth]{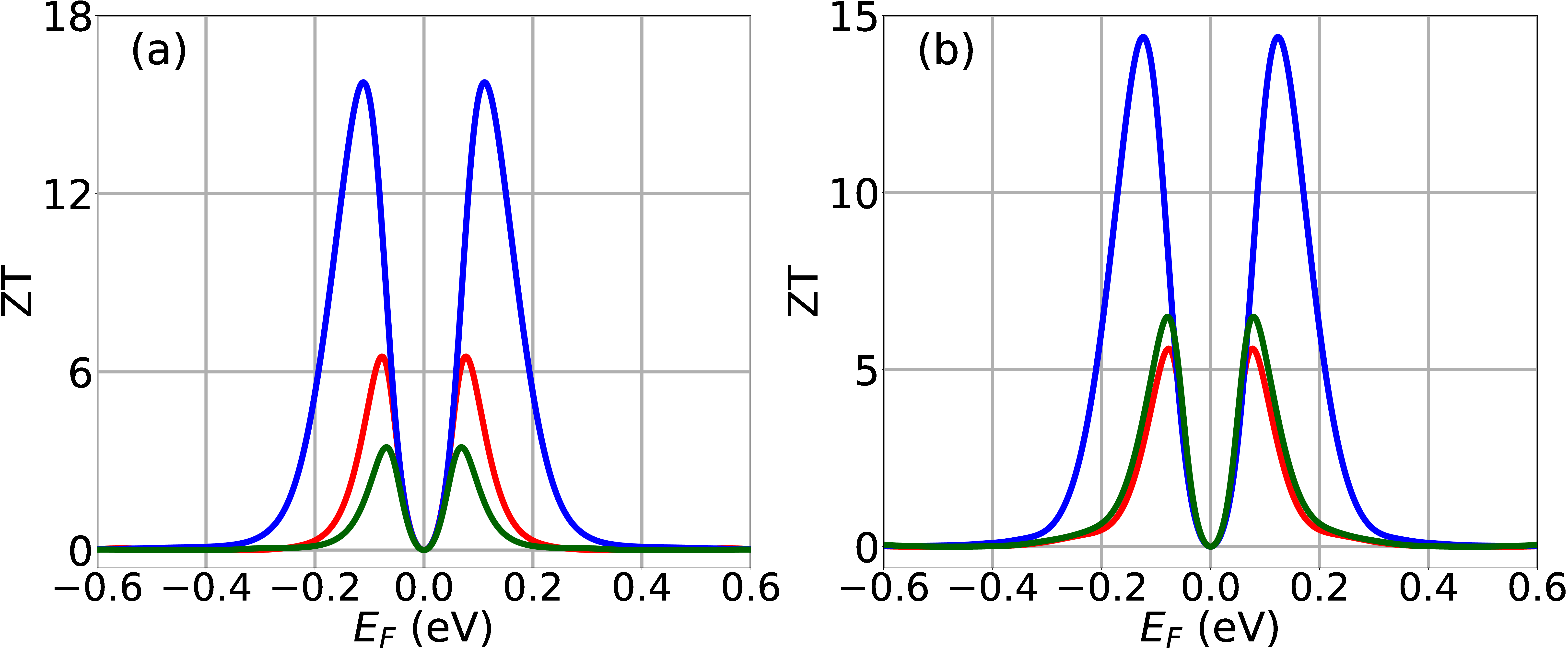}
\caption{(Color online). $ZT$ as a function of Fermi energy $E_F$ for different light parameters. (a) $AA$-stacked BLG and (b) $AB$-stacked BLG with zigzag edges. The light parameters are $A_x =1$, $A_y = 0.3$, $\varphi=\pi$, $A_x =0.5$, $A_y = 0.2$, $\varphi=\pi/2$, and $A_x =1$, $A_y = 0.3$, $\varphi=\pi/6$. The corresponding results are denoted with red, blue, green colors, respectively. The temperature is fixed at $T=300\,$K. The length and width of the system are fixed at 14$\,$nm and 14.2$\,$nm, respectively.}
\label{s2}
\end{figure}

\end{document}